\documentclass[10pt,conference]{IEEEtran}

\IEEEoverridecommandlockouts

\usepackage{graphicx}
\usepackage{textcomp}
\usepackage{makecell}
\usepackage[bottom]{footmisc}
\usepackage{xcolor}
\usepackage{multicol}
\usepackage{multirow}
\usepackage{balance}       % to better equalize the last page
\usepackage{graphics}      % for EPS, load graphicx instead 
\usepackage[T1]{fontenc}   % for umlauts and other diaeresis
\usepackage{booktabs}
\PassOptionsToPackage{hyphens}{url}
\usepackage{microtype}        % Improved Tracking and Kerning
\usepackage{ccicons}          % Cite your images correctly!
\usepackage{url}
\usepackage{cite}

\usepackage[ampersand]{easylist}
\usepackage{tabularx}
\usepackage{rotating}
\usepackage[many]{tcolorbox}
\usepackage{soul}

\usepackage{colortbl}
\usepackage{verbatim}
\usepackage{array}
\usepackage[inline]{enumitem}
\usepackage[export]{adjustbox}
\usepackage{pifont}
\usepackage{hyperref}

\usepackage{booktabs}   % For enhanced table lines
\usepackage{graphicx}   % For the \resizebox command
\usepackage{float}      % For using [H] placement specifier

\usepackage{booktabs}   % For enhanced table lines
\usepackage{graphicx}   % For the \resizebox command
\newif\ifdraft
\draftfalse %hides boldifications and comments
\makeatother

\usepackage{tikz}
\newcommand{\circlednum}[1]{\tikz[baseline=(char.base)]{
    \node[shape=circle,draw=black,fill=black,text=white,inner sep=1pt] (char) {\small #1};}}

\begin{document}

%\def\BibTeX{{\rm B\kern-.05em{\sc i\kern-.025em b}\kern-.08em
%    T\kern-.1667em\lower.7ex\hbox{E}\kern-.125emX}}

%__________________ end packages and commands from paper__________

\title{\fontsize{19pt}{24pt}\selectfont Design, Implementation and Evaluation of a Novel Programming Language Topic Classification Workflow}

%\IEEEauthorblockN{Mariam Guizani}
%\IEEEauthorblockA{\textit{Queen's University} \\
%Kingston, Canada \\
%mariam.guizani@queensu.ca}}

   \author{
    \IEEEauthorblockN{Michael Zhang}
    \IEEEauthorblockA{
        \textit{Queen's University} \\
        Kingston, Canada \\
        18jz111@queensu.ca
    }
    \and
    \IEEEauthorblockN{Yuan Tian} % Replace with the actual name
    \IEEEauthorblockA{
        \textit{Queen's University} \\
        Kingston, Canada \\
        y.tian@queensu.ca
    }
    \and
    \IEEEauthorblockN{Mariam Guizani}
    \IEEEauthorblockA{
        \textit{Queen's University} \\
        Kingston, Canada \\
        mariam.guizani@queensu.ca
    }
}

\maketitle

\begin{abstract}
As software systems grow in scale and complexity, understanding the distribution of programming language topics within source code becomes increasingly important for guiding technical decisions, improving onboarding, and informing tooling and education. This paper presents the design, implementation, and evaluation of a novel programming language topic classification workflow. Our approach combines a multi-label Support Vector Machine (SVM) with a sliding window and voting strategy to enable fine-grained localization of core language concepts such as operator overloading, virtual functions, inheritance, and templates. Trained on the IBM Project CodeNet dataset, our model achieves an average F1 score of 0.90 across topics and 0.75 in code-topic highlight. Our findings contribute empirical insights and a reusable pipeline for researchers and practitioners interested in code analysis and data-driven software engineering.

%identifying the file 0.9 F1 score
%up to for highlight 

%into as a new task. 
%Highlighting why needed... 
%PL ecosystem to know popular and less popular features.... how topics are used in ecosystem... frequency... stats...

\end{abstract}

\begin{IEEEkeywords}
Programming languages, Code topic classification, Code topic highlight, Code analysis 
\end{IEEEkeywords}

\section{Introduction}
Codebases often integrate a variety of language constructs to meet diverse functional and performance requirements. As a result, understanding their internal composition, particularly the distribution of language specific topics (such as core programming constructs) is essential. \textit{Code topic classification}, the task of assigning semantic labels to code fragments, offers a promising solution to this challenge. By automatically identifying language topics such as inheritance, operator overloading, or exception handling, this approach enables developers to locate relevant logic at a micro level while offering managers and analysts a macro-level view of a system’s functional landscape. 

%\yuan{i wonder if we need a citation here, do existing studies define code topic classification task? i think they do, but they did not apply to language constructs?}. \michael{There are code topic classification task but mostly in the high level region.  I have added this to the related work. But I don`t see one that is related to language construc } \yuan{ok, then we just define it here}

%\yuan{i removed some text here, please check latex source if they are needed}%For development teams, educators, and tool builders, identifying and localizing core programming constructs and analyzing their distribution can support more informed technical decision-making, improves developer onboarding, and enhances tooling and curriculum design.

Practical applications of topic classification are numerous. \circlednum{1} In industry and open-source contexts, onboarding new developers is often slowed by the steep learning curve of unfamiliar codebases \cite{9401978}. Especially in open-source projects, many technical challenges stem from code complexity and contributors' varying levels of prior knowledge \cite{guizani2021long}.  Identifying high-frequency topics can help developers prioritize their learning and reduce ramp-up time \cite{noauthor_choosing_2014,noauthor_how_2021} and retain contributors \cite{guizani2022attracting}. \circlednum{2} In code translation, topic classification can ensure structural and semantic fidelity. For instance, when translating a C++ abstract class structure into Python, recognizing the code's topic (i.e., abstract classes) guides the translator to correctly utilize Python’s \textsc{ABC} module or inheritance mechanisms, effectively retaining the intended polymorphic behavior. \circlednum{3} Beyond comprehension and translation, topic distribution analysis informs strategic decision-making. Teams may choose to invest in widely used constructs or deprecate niche features that incur disproportionate maintenance costs. Comparing topic prevalence across related projects or domains can also reveal domain-specific practices, such as emphasizing concurrency in real-time systems or cryptographic patterns in security-critical software. \circlednum{4} Finally, for researchers and educators, systematic analysis of topic distributions across repositories or communities can uncover changes in programming paradigms, track the adoption of advanced or new language constructs, and guide curriculum development \cite{sadowski2011heuristic,sheiner_user_2021}. As software systems grow and diversify, automated topic classification will become increasingly vital for sustaining clarity, adaptability, and informed decision-making in code-centric environments.

In this paper, we present the design, implementation, and evaluation of a novel programming language topic classification workflow. As far as we know, no previous study has defined or evaluated code-topic classification at the level of concrete programming-language constructs, such as \texttt{Template}, \texttt{Operator overload} in C++.  Prior research on semantic code classification typically focuses on high-level functional or domain-specific concepts \cite{Plösch1}\cite{ErikLinstead}\cite{DBLP:journals/corr/abs-1906-01032}. \textbf{Our aim is to build a foundation for generalized code-topic classification that enables the automatic identification and highlighting of these constructs. By framing the task at token level, our work is the first to fill this fine-grained gap.} %\yuan{we should explicitly mention our novelty here, e.g., As far as we know... depending on the answer to by previous question early this section}. \michael{Please see the above edited section}

The proposed workflow combines statistical feature extraction using TF-IDF with a multi-label Support Vector Machine (SVM) classifier \cite{cortes1995support}, and introduces a post-processing pipeline that incorporates a character-level sliding window, voting mechanism, and syntactic boundary expansion. Through evaluation on the CodeNet dataset \cite{puri_codenet:_2021}, we demonstrate the effectiveness of our approach in accurately identifying and highlighting code topics such as operator overloading, templates, and virtual functions in the context of C++. Our findings offer empirical insights and a reusable, extensible pipeline for researchers and practitioners seeking to integrate code topic classification into tools for software comprehension, translation, education, and maintenance.

\section{Methodology}

%As shown in Figure 1, in this study, we first start by selecting a programming language and topics that will be used in this study. Later, we filtered out the code samples related to the target topic from the Project Codenet dataset and created our own new dataset. Then, we improved the data quality by removing comments and redundant blanks and balanced the distribution of samples for each topic using data augmentation methods. Next, TF-IDF is used to convert code fragments into high-dimensional vectors to capture key statistical patterns related to syntax and structure, and a multi-label SVM (One-vs-Rest) model is utilized to identify multiple topics simultaneously. After evaluating the accuracy, precision, recall, and F1 score of the model through cross-validation, we perform post-processing in the inference stage by combining sliding window, voting mechanism,  and ultimately output code snippets with highlighted annotations, which facilitates the precise localization and visualization of multiple C++ topics. 

\begin{figure}[H]
    \centering
    \includegraphics[width=\linewidth]{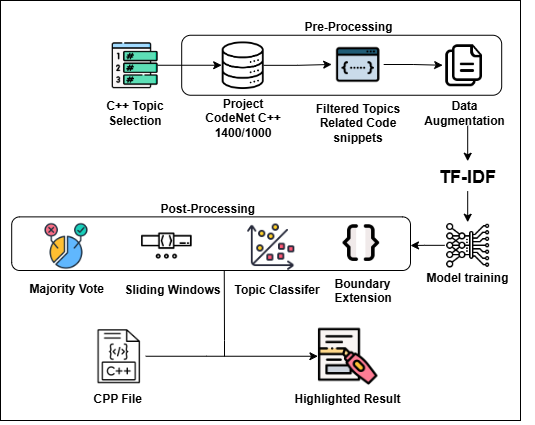} 
    \caption{Overview of the proposed workflow for automatic classification of (language construct) topics.}
    \label{fig:overview}
\end{figure}

Figure~\ref{fig:overview} presents an overview of our proposed workflow, which comprises five key stages: (1) selecting the target language and the corresponding topic set; (2) constructing a high-quality labeled dataset; (3) training a multilabel classifier using TF-IDF features and an SVM model; (4) applying a post-processing pipeline to allow fine-grained topic localization (highlighting); and (5) evaluating the effectiveness of the code topic highlighting workflow.

 %from the CodeNet dataset \cite{puri_codenet:_2021} \yuan{this is technical details?}

\subsection{Language and topic selection}
We selected C++ as the focus programming language for this study due to its enduring popularity and practical relevance. It ranks 6\textsuperscript{th} on GitHub, accounting for 10\% of public contributions~\cite{octoverse2022}, and has consistently held a top 2 position in the TIOBE index as of June 2025~\cite{tiobe2025rank}. 

We then selected a set of advanced C++ language topics from an authoritative reference~\cite{cppreference}, namely ``\textsc{Operator overload}'', ``\textsc{Virtual functions}'', ``\textsc{Friend functions}'', ``\textsc{Inheritance}'', ``\textsc{Inline functions}'', ``\textsc{Templates}'', ``\textsc{Classes}'', and ``\textsc{Try-catch}''. These topics were chosen based on their pedagogical importance~\cite{10.5555/557014}, prevalence in real-world software systems, and frequent occurrence in open-source repositories. Furthermore, their long-standing support and evolution across multiple C++ standards make them both stable and representative, offering a strong foundation for code topic classification research.

%We selected C++ due to its widespread usage, ranking 6th on GitHub with 10\% of public contributions \cite{octoverse2022}, and consistently maintaining a top 2 position in the TIOBE index from June 2024 \cite{tiobe2024}. 
%We selected a group of advanced C++ language topics from the authoritative resource \cite{cppreference} as our research topics, namely ``\textsc{operator overloads}'', ``\textsc{virtual functions}'', ``\textsc{friend functions}'', ``\textsc{inheritance}'', ``\textsc{Inline functions}'', ``\textsc{recursion}'', ``\textsc{templates}'', ``\textsc{classes}'', ``\textsc{function pointers}'', and ``\textsc{try-catch}''. We chose these specific topics because these topics are widely used in teaching environments \cite{10.5555/557014} and in real-world development. In addition, these topics can be easily recognized in numerous instances of large open-source projects, thus supporting data collection and model training with ample heterogeneous resources. Moreover, these topics have been stabilized or enhanced in different versions of C++, making them robust in generality and representativeness, thus providing a solid foundation for subsequent in-depth research.

\subsection{Dataset creation and pre-processing} \label{sec:preprocess}

In this study, we use two subsets of the CodeNet dataset~\cite{puri_codenet:_2021}: $\text{Project\_CodeNet\_C++1400}$ (420{,}000 \texttt{.cpp} files) for model training and $\text{Project\_CodeNet\_C++1000}$ (500{,}000 \texttt{.cpp} files) for evaluating highlight quality. CodeNet is a widely used benchmark dataset contributed by IBM, containing practical programs written in 55 programming languages, including C++, sourced from two competitive programming platforms: AIZU Online Judge and AtCoder. It is commonly used to benchmark algorithmic performance, especially for automatic code generation and translation \cite{macedo2025intertrans}. Compared to typical open-source repositories, Project CodeNet submissions are generally shorter (tens to a few hundred lines), algorithm-focused, and consistently structured, offering the scale and quality that make the dataset particularly suitable for machine learning research on code.

%We leverage Project CodeNet\cite{puri_codenet:_2021} to train/ test our model  training/ testing dataset for our project.  The data is drawn from two renowned competitive programming platforms, AIZU Online Judge and AtCoder, which are used to benchmark the performance of C++ code. In this study, we selected two subsets of IBM's Project CodeNet: $\text{Project\_CodeNet\_C++1000}$ (including 500,000 \texttt{.cpp} files) and $\text{Project\_CodeNet\_C++1400}$ (including 420,000 \texttt{.cpp} files). Both subsets are organized into multiple directories and each is the solution for a specific problem, offering a rich set of coding problems, problem‐solving approaches, and individual coding styles. The reason we chose Project CodeNet was because of its combination of high quality and simplicity, making it particularly suitable for machine learning tasks such as code classification and similarity detection. Its large size ensures that there is sufficient training data for any chosen topic. In our project, we will be using $\text{Project\_CodeNet\_C++1400}$ for model training and $\text{Project\_CodeNet\_C++1000}$ for evaluating the quality of highlights.

We used a rule-based heuristic approach to extract topic-related code snippets and construct our ground truth dataset (see supplemental material~\cite{zhang2025supplementary}). Each ground truth example is a standalone code snippet (i.e., a file) containing a single, heuristically identified instance of a target language construct topic, extracted from raw \texttt{.cpp} files. To mitigate class imbalance across topics, we applied lightweight data augmentation techniques~\cite{noauthor_what_2024} that preserve code semantics while increasing sample diversity. This augmentation strategy enables us to expand underrepresented classes without compromising the quality of the dataset. It is important to note that the actual evaluation of our workflow is conducted on raw .cpp files rather than on these preprocessed training snippets (see Section~\ref{sec:evaluation}).

%\mariam{this reference is not showing do we still need it}(the imbalance nature could be observed from the total data column in Table \ref{Table 2:classification_report})

To further validate our rule-based heuristic, we manually annotated a random sample of 10 raw \texttt{.cpp} per topic, each containing a ground truth snippet identified by the heuristic approach. We manually highlighted the relevant topic constructs in the raw file and compared them against the labels produced by our heuristic-based approach. The results showed that the extracted snippets achieved over 90\% precision, recall, and F1-score on average at the character level compared to human-highlighted code.

\subsection{Model training} \label{sec:modeltraining}

The goal of model training in our workflow is to develop a model that can automatically identify whether a given C++ code snippet demonstrates one or more target language construct topics. This is based on the observation that many language constructs exhibit recurring patterns of token usage, which can be captured through statistical modeling.

To achieve this, we adopt a multi-label classification approach using Support Vector Machines (SVM) with a binary relevance strategy. The training data consists of preprocessed C++ source files obtained from the previous step (by applying rule-based heuristic on $\text{Project\_CodeNet\_C++1400}$). Each snippet is assigned to one target language construct. These labeled examples are then encoded into a binary matrix format using the \texttt{MultiLabelBinarizer}~\cite{noauthor_multilabelbinarizer_nodate} to support multi-label model training. Independent SVM classifiers are then trained in a one-vs-rest configuration, enabling scalable and interpretable topic-level predictions. While SVM is a classical machine learning method, it remains highly effective for high-dimensional sparse data, such as TF-IDF representations of source code, due to its strong generalization ability and robustness to overfitting. Additionally, the use of SVMs allows for efficient training and clear decision boundaries, making them suitable for establishing reliable baselines in multi-label classification tasks where interpretability and computational efficiency are important.

To extract discriminative features from source code, we apply character-level TF-IDF encoding using \texttt{TfidfVectorizer}~\cite{noauthor_tfidfvectorizer_nodate}, with an \texttt{ngram\_range} of (1, 5). This setup captures not only keywords but also syntax-critical elements such as brackets, operators, and template meta-programming artifacts that are key for robust C++ topic classification. Character-level tokenization enhances feature extraction in C++ by reliably capturing critical syntactic elements, such as overloaded operators, template brackets, and punctuation, that word- or token-level approaches often overlook.

\subsection{Post-processing (for highlighting)}
%After obtaining a reliable classification model, the next step is how to apply it effectively to C++ source code. 

While the trained SVM model enables us to determine whether a code snippet is associated with one or more language construct topics, our ultimate goal is more fine-grained: to \textbf{highlight} (i.e., locate) the specific region within the code (i.e., a sequence of tokens) that corresponds to each identified topic. Since the model operates at the file or snippet level (e.g., during testing, a full \texttt{.cpp} file is given as input), it cannot directly highlight the specific code regions relevant to each topic. To bridge this gap, we introduce a post-processing step (see Figure \ref{fig:overview}) that applies (1) a sliding window strategy for sequence labeling, (2) a voting mechanism to enhance classification robustness, and (3) a boundary expansion mechanism to align highlights with structural code regions.

The core idea of the sliding window strategy is to move a fixed-size window across the input source file and evaluate each window independently to determine whether it corresponds to any target topic (using the model trained in the previous step, ref. Section\ref{sec:modeltraining}). The predictions for individual windows are then aggregated to produce the final highlighting result. This process relies on two key parameters: the \textsc{window size}, which defines the number of characters per segment; the \textsc{step size}, which determines how many characters the window moves at each step. In this paper, we use a step size of one character to enable fine-grained and precise highlighting.

%We define a window size for each topic based on its typical code length. 
We derive topic-specific default \textsc{window size}s from experimentation to capture local context more effectively and improve classification accuracy. For example, we use a 20-character window for ``\textsc{Operator Overload}'' and a 40-character window for ``\textsc{Virtual Function}''. A complete list of window sizes for all topics is provided in the supplemental material \cite{zhang2025supplementary}. %These default topic-specific window sizes can be customized according to the specific application needs.

%In the sliding window strategy, we first have to define a window size threshold for each topic. Based on our experimentation with different window sizes, we found that using the typical topic length as a proxy for window size threshold captures local contexts more effectively and significantly improves overall accuracy.
%Based on the observations of a series of exploratory tests, we found that an appropriately chosen window size not only captures local contexts more effectively but also significantly improves overall accuracy. 
%These varied window sizes are based on the typical length of each topic in the code. 

%\yuan{removed, as already captured in early part}Our experiments show that some topics are easily identified using short context windows, while others require a broader view of surrounding code. In our approach, we slide a fixed-size window character by character across each source file and classify the content within each window independently.

%can produce unstable predictions in localized regions. Overlapping topics or partially visible patterns within a window can lead to misclassifications or missed detections. %Instead of relying on a single binary decision (0 or 1), 
However, the sliding window method alone does not specify how to integrate predictions, particularly when overlapping windows yield conflicting labels (e.g., one is classified as containing the target topic while an adjacent one is not). To resolve this, we implement a voting strategy that aggregates predictions across overlapping windows to produce a more stable and consistent highlight. 

Specifically, to determine whether a character should be highlighted for a given topic, we compute a confidence score (see Equation~\ref{eq:Equation1}) based on the proportion of sliding windows covering that character that classify it as belonging to the topic. In our trial experiments, we observed that this strategy significantly reduces false positives and increases highlight robustness. 

{
\setlength{\abovedisplayskip}{6pt}
\setlength{\belowdisplayskip}{6pt}
\setlength{\abovedisplayshortskip}{6pt}
\setlength{\belowdisplayshortskip}{6pt}
\begin{equation}
\label{eq:Equation1}
\text{Confidence}(c, topic_i) =
\frac{\text{HighlightCount}(c, topic_i)}{\text{WindowNum}(c, topic_i)},
\end{equation}
}

Equation~\ref{eq:Equation1} defines the confidence score for character \( c \) with respect to a specific topic \( topic_i \). This score quantifies the proportion of sliding windows covering character \( c \) that are predicted to contain \( topic_i \). Specifically, \(\text{HighlightCount}(c, topic_i)\) represents the number of sliding windows that both include character \( c \) and are classified as containing \( topic_i \), while \(\text{WindowNum}(c, topic_i)\) denotes the total number of windows that include character \( c \). The resulting confidence score ranges from 0 to 1 and is used to determine whether \( c \) should be highlighted for \( topic_i \), based on whether it exceeds a predefined threshold parameter, i.e., \textsc{Highlight Threshold}. Based on empirical experimentation, we found that setting the \textsc{Highlight Threshold} to 80\% consistently yields the most accurate results. This range offers a good trade-off between precision and recall, while improving the robustness of localized highlights.

Finally, we supplement the sliding window and voting strategies with a boundary expansion mechanism to address cases where highlights overshoot or undershoot the actual topic scope. This mechanism is particularly useful for topics that typically span entire functions. After the initial highlighting, we traverse the code to identify matched pairs of opening and closing braces. If a highlighted region is found to lie within a valid function boundary, we expand the highlight to encompass the entire brace-enclosed block. This step ensures that the final annotations better reflect the structural semantics of C++ code and mitigates issues such as fragmented or incomplete highlights introduced by character-level windowing.

\subsection{Visualization demo}\vspace{-0.35\baselineskip}
\begingroup
\setlength{\textfloatsep}{8pt}
\setlength{\intextsep}{8pt}
\setlength{\abovecaptionskip}{4pt}
\setlength{\belowcaptionskip}{4pt}
\begin{figure}[t]
    \centering
    \includegraphics[width=\linewidth]{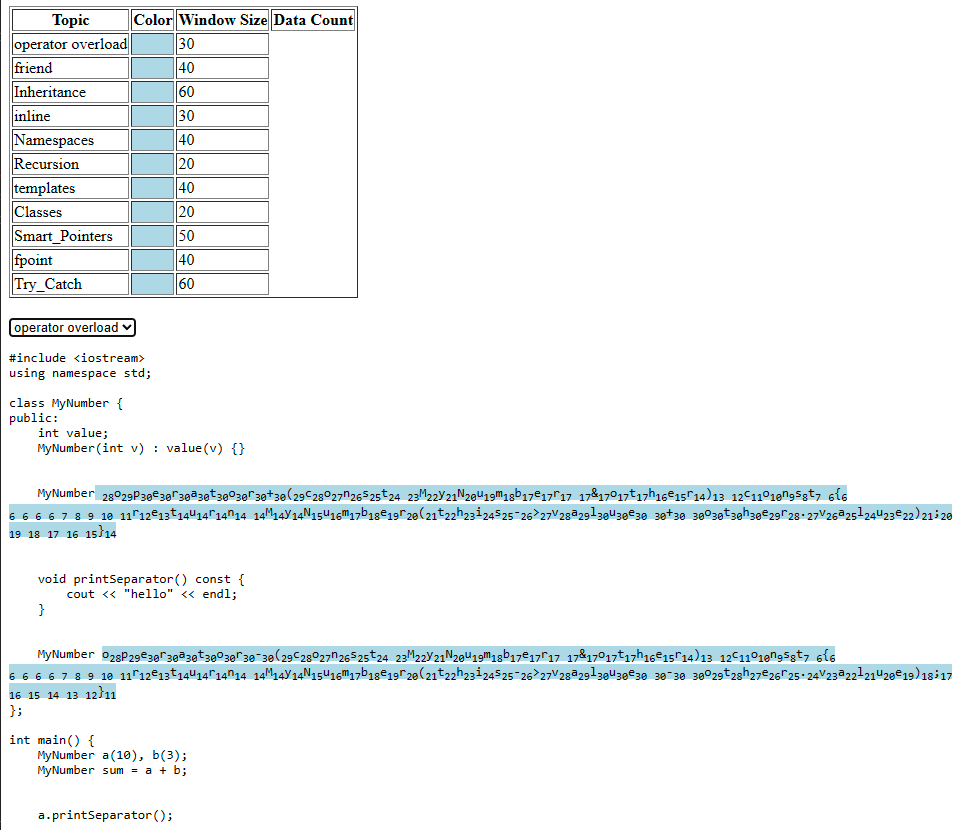}
    \caption{Visualization of the highlighting results for the \textsc{Operator Overload} topic. The table (top) shows topic-specific parameters, including window size. The code snippet (bottom) highlights character regions predicted to belong to the selected topic. Our visualization tool can be accessed at \href{https://cpp-classifier.onrender.com/}{https://cpp-classifier.onrender.com/}.}
    \label{fig:html-demo}
\end{figure}
\endgroup

To support intuitive exploration and experimentation, we implemented a visualization of our workflow results, as shown in Figure~\ref{fig:html-demo}.  A link to the hosted  visualization is provided in the supplementary material~\cite{zhang2025supplementary}.

As shown in the demo figure, a drop-down menu is positioned at the top left part of the HTML page, allowing users to select a target topic. As shown in the demonstration, the current topic selected is \textsc{"Operator Overload"}. In the code section, two operator overload functions are correctly highlighted, indicating the effectiveness of our implementation. 

\subsection{Evaluation} \label{sec:evaluation} 
To evaluate the performance and reliability of our workflow, we measure (1) the performance of the classifier on the training data, i.e., \texttt{Project\_CodeNet\_C++1400}, and (2) the performance of the overall highlight workflow on the testing data, i.e., \texttt{Project\_CodeNet\_C++1000}. The former assesses how well the trained model distinguishes language construct topics based on labeled snippets, while the latter evaluates the effectiveness of the full pipeline, including sliding window, applying the classifier, voting, and boundary expansion, in producing accurate highlights on raw source files.

To evaluate the classifier's performance, we utilize the most widely used classification metrics: precision, recall, and F1-score. We train and test the model using 10-fold cross-validation with \texttt{MultilabelStratifiedKFold}~\cite{trent-b_trent-b/iterative-stratification_2025}, ensuring balanced topic distribution across folds. Precision, recall, and F1-score are computed using \texttt{classification\_report} from \texttt{sklearn\.metrics}, and final performance is reported as the average across all folds. Since the classifier operates on individual code snippet files, all evaluation is performed at the file level.

To assess the performance of the highlighting workflow, we calculate precision, recall, and F1-score at the character level by comparing the highlights generated by our workflow with the ground truth annotations (derived via rule-based heuristics, see Section~\ref{sec:preprocess}). A character is considered a true positive if it is correctly highlighted for a given topic in both the model output and the ground truth. Precision is then computed as the proportion of correctly highlighted characters (true positives) out of all characters predicted as highlighted by the model. For example, if the model highlights 100 characters for a topic and 80 of them match the ground truth, the precision is 80\%. Recall is calculated as the proportion of correctly highlighted characters out of all characters that should have been highlighted according to the ground truth.

Given that only a few percent of source files contain a target topic, to ensure a fair and tractable evaluation across topic classes, we apply the highlighting workflow to a filtered subset of files that are known, based on the ground truth, to contain the target topic. For topic classes with fewer than 300 files (containing the topic), all available examples are used; for more frequent classes, we randomly sample 300 examples. This setup enables us to focus the evaluation on fine-grained localization performance, assuming the topic is present, thereby complementing the file-level classification evaluation.

\section{Results}

In this section, we discuss the results for the two experiments described in Section \ref{sec:evaluation}.

\subsection{Results of classifier}

Table \ref{tab:classification_report} presents the performance of our topic classifier for each topic, with each row representing a specific topic and the columns showing precision, recall, F1-score, support, and total data. We define ``support'' as the number of samples used per fold, and ``total data'' as the total number of training samples per topic. The ``Average'' row at the bottom reports aggregate performance across all classes.

\begin{table}[htbp]
  \centering
   \caption{Per-topic precision, recall, and F1-score of the classifier. ``Support'' indicates the number of samples in a single fold; ``Total data'' refers to the total number of samples used for training across all folds.}
  \resizebox{0.5\textwidth}{!}{
      \begin{tabular}{l c c c c c}
        \toprule
        \textbf{Class} & \textbf{Precision} & \textbf{Recall} & \textbf{F1-Score} & \textbf{Support} & \textbf{Total Data} \\
        \midrule
        Classes           & 0.95 & 0.96 & 0.95 & 97  & 970 \\
        Inheritance       & 1.00 & 0.98 & 0.99 & 204 & 2040 \\
        Namespaces        & 1.00 & 0.99 & 0.99 & 261 & 2610 \\
        Try\_Catch        & 1.00 & 1.00 & 1.00 & 138 & 1381 \\
        Friend            & 1.00 & 0.99 & 1.00 & 114 & 1142 \\
        Inline            & 0.73 & 0.96 & 0.83 & 593 & 5930 \\
        Operator overload & 1.00 & 0.99 & 1.00 & 494 & 5439 \\
        Templates         & 0.94 & 0.58 & 0.71 & 495 & 4956 \\
        Virtual function  & 1.00 & 1.00 & 1.00 & 62  & 621 \\
        \midrule
        Average       & 0.90 & 0.90 & 0.90 & 2524 & -- \\
        \bottomrule
        \\
        
      \end{tabular}
    }
 
  \label{tab:classification_report}
\end{table}

The results indicate that the classifier performs consistently well with an average precision, recall, and F1-score of 0.9. For most topics, the precision and recall values are equal to or near 1.00. Exceptions include ``\textsc{Inline}'' and ``\textsc{Template}'', which show reduced precision and F1-scores. For example, ``\textsc{Inline}'' achieves a high recall of 0.96 but a lower precision of 0.73, resulting in an F1-score of 0.83.

\begin{figure}[h]
\centering
\includegraphics[width=0.5\textwidth]{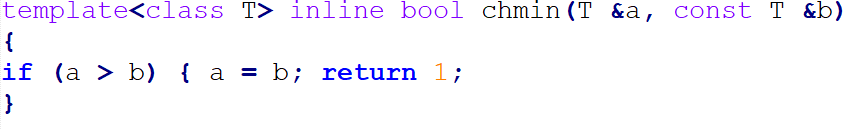} 
\caption{Example of template and inline co-occurrence.}
\label{fig:template}
\end{figure}

Upon inspection, these lower scores are largely attributable to topic co-occurrence. In particular, ``\textsc{Inline}'' and ``\textsc{Template}'' can appear within the same code file (Figure~\ref{fig:template}), making it difficult for the model to disambiguate them. This is a known challenge in multi-label classification with binary relevance strategies, where overlapping features may result in mutual suppression between classifiers. Additionally, TF-IDF representations, while lightweight and effective for most topics, do not explicitly encode structural relationships, which may further limit discriminative power when features from multiple topics are intertwined.

Despite these challenges, the model demonstrates strong performance across the remaining topics, such as ``\textsc{Classes}'', ``\textsc{Try-Catch}'', ``\textsc{Friend Function}'', and ``\textsc{Operator Overload}'', all of which report F1-scores nearing 1.00. These topics are often associated with distinctive patterns that are reliably captured by character-level TF-IDF, and they exhibit minimal semantic overlap.

\subsection{Results of highlight}

%While the classification evaluation looks at file level labeling our code topic highlight evaluation investigates character-level classification. 
%\mariam{This reference is not showing, could you please rectify at Michael} 
Table \ref{tab:average_f1} displays the precision, recall, and F1-score of our code topic highlight. On average, the code-topic highlight achieves a precision of 0.81, a recall of 0.84, and an F1-score of 0.75. %\yuan{why not add support for this table? we haven't reported how many data per topic in the testing dataset, I assume table 1 is from training dataset only by doing cross validation, table 2 is from testing dataset, so table 1's support does not apply to table 2} \michael{added support in table 2. will also mention in text of Evulation section}

\begin{table}[htbp]
  \centering
  \caption{Per-topic precision, recall, F1-score and support of the highlight (measured at character-level).}
  \begin{tabular}{lcccc}  
\toprule
\textbf{Class} & \textbf{Precision} & \textbf{Recall} & \textbf{F1-Score} & \textbf{Support}\\
\midrule
Classes            & 0.87 & 0.78 & 0.80 & 300\\
Inheritance        & 0.91 & 0.94 & 0.91 & 34\\
Try\_Catch         & 0.99 & 0.70 & 0.79 & 258\\
Friend             & 0.93 & 0.99 & 0.96 & 107\\
Inline             & 0.25 & 0.99 & 0.37 & 300\\
Operator overload  & 0.86 & 0.99 & 0.90 & 300\\
Templates          & 0.86 & 0.98 & 0.89 & 300\\
Virtual function   & 0.95 & 0.99 & 0.96 & 19\\
\midrule
\textbf{Average} & 0.81 & 0.84 & 0.75 & --\\
\bottomrule
\end{tabular}
   \label{tab:average_f1}
\end{table}

In general, most topic achieve an F1-score of 0.80 or higher. This indicates that our post-processing stage (sliding window, voting mechanism and boundary expansion) is both functional and largely effective. 

%\mariam{@Michael please update the numbers in text}\yuan{now I see classes has a F1 of 0.8 rather than 0.54? wonder if the average is still accurate, please double check this table 2} \michael{I have rerun the Classes and virtual function and both got much improved. Inline stays the same. The result in the table is the accurate number. I did not update the number in the text. Now should be all good.}

However, we do notice a lower highlight performance for  ``\textsc{Inline}'', with an F1-score of 0.37. When it comes to inline functions, the model achieves near-perfect recall (0.99) but very low precision (0.25), indicating overprediction. To investigate, we tested the ``\textsc{Inline}'' topic alone and found that many non-inline sections of the code were still being highlighted. Our hypothesis is that the model is overly sensitive to this topic: tokens such as \verb|define|, which merely contain the substring “ine”, are frequently misclassified as ``\textsc{Inline}''. The subsequent bracket-expansion step amplifies this problem, extending the highlight over large blocks of code. As a result, the entire scope is highlighted as \verb|inline|, increasing recall and hindering precision down.

\section{Related Work}

Existing work on code topic classification has shown that machine learning techniques can effectively uncover high-level concerns in source code, thereby facilitating program comprehension. A common approach involves applying Latent Dirichlet Allocation (LDA), a popular automated topic modeling technique, to entire source code files. This method has been used to identify topics such as GUI and persistence, and to analyze how these concerns are distributed across a software system~\cite{ErikLinstead}. Building on this idea, Plösch et al. introduced the Technical Topic Classification (TTC) model, which maps static analysis findings to a comprehensive catalog of technology-related topics and aligns them with ISO-9126 software quality attributes~\cite{Plösch1}. More recently, Gelman et al. proposed a language-agnostic, character-level neural model that predicts a diverse set of semantic tags for code snippets. Their approach operates at the function level and achieves strong generalization across programming languages~\cite{arxiv}. Alreshedy et al. proposed SCC, a machine learning tool for classifying the programming language of short code snippets extracted from Stack Overflow posts. Their approach employs a Multinomial Naive Bayes classifier trained on TF-IDF features. \cite{alreshedy2018sccautomaticclassificationcode}

%Code topic classification has proven valuable in decision support for software evolution. Kim et al. \cite{4408585} used an SVM classifier to predict whether code changes would introduce errors with an accuracy comparable to existing methods and significantly reduced error detection time. Similarly, Arcelli Fontana et al. achieved high accuracy using SVM and other classifiers to detect code odors (e.g., ``god classes''), thus providing objective and flexible support for software refactoring decisions. Together, these studies emphasize the effectiveness of applying machine learning-based code classification techniques to improve software quality assurance processes.

 However, prior work has primarily focused on high-level functional or domain-related topics (e.g., middleware). While SCC proposed by Alreshedy demonstrates the feasibility of snippet-level classification, its focus is limited to language identification rather than fine-grained programming language constructs. In addition, these approaches typically operate at the file level, lacking the resolution needed to identify and localize specific topics within individual lines and across contiguous segments of code. In contrast, we formulate a new problem of fine-grained code topic classification and present an automated workflow to address it. Our solution is evaluated on C++ program constructs and demonstrates promising performance.

\section{Future work}
As the first study on fine-grained code topic classification, our work faces several threats to validity, which we aim to address in future research. 

First, the accuracy and robustness of our C++ topic classifier can be further enhanced. Our current use of a fixed-size sliding window may not fully capture locality-dependent syntactic and structural variations. Future research will explore adaptive windowing mechanisms that dynamically adjust window sizes based on code context, allowing the model to extract relevant features more precisely.

Second, our approach faces challenges in distinguishing closely co-occurring topics. To address this, future work will explore advanced multi-label classification strategies beyond binary relevance, such as classifier chains or problem transformation methods. These techniques allow better modeling of topic dependencies and could improve the model’s ability to handle overlapping labels. In particular, we plan to refine the classification of the inline topic, which currently suffers from low recall. Many misclassified examples include preprocessor directives (e.g., \#define, \#include) or macros containing the substring "inline." To improve precision and recall, we will curate a set of negative examples under a new “non-inline” class, helping the classifier draw clearer boundaries between true and false positives.

Third, we aim to extend our workflow to support additional languages and topics. While our current design is language-agnostic, i.e., we treat source code as plain text, adapting it to other languages such as Java or Python primarily involves retraining the SVM model, adjusting window sizes, and updating topic-specific regex patterns. However, the success of generalization depends on the nature of the target topics. For instance, our classifier struggles with the recursion topic due to its lack of distinctive lexical tokens. Recursive and non-recursive functions often share similar surface vocabulary, making them hard to distinguish using TF-IDF alone.

In terms of model design and input representation, several enhancements could be explored. For instance, incorporating structural information from Abstract Syntax Trees (ASTs) or leveraging advanced large language models (LLMs) may further improve performance. LLMs, such as GPT-style models, have shown strong capabilities in understanding and generating code across languages and tasks. However, they often require substantial computational resources, large-scale labeled data, and complex infrastructure, which may not be practical for all use cases. While ASTs can effectively capture structural code features, they rely on language-specific parsers, which increase system complexity and reduce portability. In contrast, our current TF-IDF-based approach is lightweight, language-agnostic, and easy to implement, making it well-suited for integration into multi-language development environments. To balance these trade-offs, we plan to augment our workflow with a minimal AST-based analysis step that extracts key structural elements, such as recursive function calls, loops, or conditional constructs. This hybrid strategy aims to improve classification accuracy for structure-dependent topics while preserving the simplicity and generalizability of our original design.

\section{Conclusions}
In this study, at its core, the workflow combines multi-label classification with post-processing to enable fine-grained localization of C++ topics such as classes, inheritance, and try-catch. During the evaluation phase, we conducted visual highlighting tests by comparing the automatically generated highlights with manually annotated ground-truth regions. We evaluated our model using 10-fold cross-validation and measured its performance with standard classification metrics, including precision, recall, and F1-score. The model consistently achieved F1-scores above 90\% for most topics, demonstrating strong classification capability. To assess the effectiveness of the highlighting workflow, we further evaluated its performance on a held-out dataset using character-level precision, recall, and F1-score. The workflow achieved an average F1-score of 75\%, indicating its potential for accurately localizing fine-grained code topics.
%RHowever, certain closely related features, such as ``\textsc{Inline}'', still experienced classification errors due to its frequent co-occurrence and overlap. 

%This suggests that future research should incorporate more flexible multi-label classification strategies or replace TF-IDF with AST for feature extrication to further enhance the model’s accuracy and robustness. 

Overall, our workflow constitutes a proof of concept for fine‑grained, topic‑aware code analysis that can serve as a building block for downstream tasks such as automated source‑to‑source translation, accelerated program comprehension, and seamless developer onboarding.

%`` ''

% BALANCE COLUMNS
%balance{}
\bibliographystyle{IEEEtran}
\bibliography{biblio}
\end{document}